\documentclass[conference]{IEEEtran}
\usepackage{cite}
\usepackage{amsmath,amssymb,amsfonts}
\usepackage{algorithmic}
\usepackage{graphicx}
\usepackage{textcomp}
\usepackage{xcolor}

\ifCLASSINFOpdf
\else
\fi
\begin{document}
%
\title{Data Augmentation and CNN Classification For Automatic COVID-19 Diagnosis From CT-Scan Images On Small Dataset}


\author{\IEEEauthorblockN{Weijun Tan}
\IEEEauthorblockA{LinkSprite Technologies\\
Longmont, CO 80501, USA\\
Email: weijun.tan@linksprite.com}
\and
\IEEEauthorblockN{Hongwei Guo}
\IEEEauthorblockA{Shenzhen Deepcam Information Technologies\\
Shenzhen, China\\
Email: hongwei.guo@deepcam.com}
}


%


\maketitle

\begin{abstract}
We present an automatic COVID1-19 diagnosis framework from lung CT images. The focus is on signal processing and classification on small datasets with efforts putting into exploring data preparation and augmentation to improve the generalization capability of the 2D CNN classification models. We propose a unique and effective data augmentation method using multiple Hounsfield Unit (HU) normalization windows. In addition, the original slice image is cropped to exclude background, and a filter is applied to filter out closed-lung images. For the classification network, we choose to use 2D Densenet and Xception with the feature pyramid network (FPN). To further improve the classification accuracy, an ensemble of multiple CNN models and HU windows is used. On the training/validation dataset, we achieve a patient classification accuracy of 93.39\%.  
\end{abstract}


%
\IEEEpeerreviewmaketitle

\section{Introduction}

The outbreak of the novel COVID-19 coronavirus in late 2019 has put a tremendous threat to the whole world and become one of the worst disaster in the human history. As of end late April 2021, more than 142 millions infections have been identified, more than 3 million lives have been lost, and more than 200 countries have been drastically overwhelmed \cite{WHO}. Therefore, it is very critical to stop the spreading of the virus. After a person is confirmed to have COVID-19, safety measures and treatments can be taken accordingly. In \cite{COVID-trace}, thermal imaging is used to detect fever patients and face recognition is used to report and trace patients and their close contacts. Among the techniques to diagnosis COVID-19, X-ray and CT-scan images are studied extensively.

In this paper, we present an automatic diagnosis framework from chest CT scan images. Our goal is to classify COVID-19, Community-Acquired Pneumonia (CAP), and normal cases from a volume of CT scan images of a patient. We use the dataset provided in the Signal Processing Grand Challenge (SPGC) on COVID-19 of the IEEE ICASSP 2021 \cite{SPGC}. Our preliminary study shows that one major challenge is that the training/validation dataset is small. This challenge is common to many other datasets, and different approaches are studies to address this problem, or the broader across-domain dataset problem, through data augmentation, across-domain
adaptation \cite{Hybrid-COVID}, \cite{SODA}, \cite{Contrative-CovidNet}, or using the capsule network \cite{Covid-caps}, \cite{COVID-FACT}.

In this paper, we propose a novel data augmentation technique using multiple Houndsfeld Unit (HU) normalization windows. This data augmentation aims at improving a CNN model’s generalization capability. In addition, it can exploit large COVID-19 CT scan datasets that are available online but without preprocessing details. Other signal processing techniques we use include cropping the chest images to exclude background, and filtering out close-lung images.

For the classification network, after exploring 3D CNN classification networks, lung mask segmentation networks, and quite a few 2D CNN classification networks, we choose to use Densenet \cite{DenseNet} and Xception \cite{Xception} 2D CNN classification networks with the feature pyramid network (FPN) \cite{FPN}. To further improve the classification accuracy, an ensemble of multiple CNN models is used. On the provided training/validation dataset, we achieve a patient classification accuracy of 93.39\%.

\section{Related Work}

Due to the urgency of control of spreading of the COVID-19 virus, a lot of researches have been done to diagnose it using deep learning approaches, mostly CNNs on CT scan images or X-ray images. A few examples are \cite{Hybrid-COVID}, \cite{SODA}, \cite{Contrative-CovidNet}, \cite{example1}, \cite{example2}, \cite{Covid-caps}, \cite{COVID-FACT}. For a complete review, please refer to \cite{Review1} and \cite{Review2}.

These methods can be categorized to 2D, 2D+1D, and 3D based on how information from multiple slice images are aggregated and how the final decision is made. In the 2D method, a 2D CNN classification network in used to make a prediction on slice image individually. Then to make a decision for a patient, some voting method is typically used \cite{SPGC1}, \cite{COVID-CTSet}, \cite{COVID-FACT}. Others use a 2D CNN network on slice image to generate embedding feature vector for every image, then the feature vectors of selected multiple or all slice images are pooled to a single global feature vector, and finally a small classification network (typically just a few fully-connection (FC) layers) is used to make a final decision. This is called 2D+1D method \cite{li2020artificial}, \cite{CTCAPS}, \cite{SPGC2}. In these two methods, annotation for slice image is needed. The third method is a pure 3D CNN network, where slice annotation is not needed, and a selected set of or all the slice images are used as input, and the 3D network process all these input images all at once in a 3D channel space \cite{Tongji}, \cite{SPGC3}, \cite{He2021CovidNet3D}. 

In the 2D CNN method, some use the lung mask segmentation, but most of them directly use the raw slice image.  The COVID-MaskNet \cite{COVID-CT-Mask-Net} uses a segmentation network to localize the disease lesion, then use a FasterCNN-based approach to do the classification on the detected lesion regions. The COVID-Net Initiative \cite{Gunraj2020}, \cite{Gunraj2021} have done extensive studies of COVID classification on both CT scan images and X-ray images. They also collect and publish the largest CT image dataset - so called COVIDx CT-2 dataset. In \cite{COVID-CTSet}, Resnet50 with FPN is used. In \cite{SPGC1}, a combination of infection/non-infection classifier, and a COVID-19/CAP/normal classifier is used. 

In the 2D+1D method, in \cite{li2020artificial}, a pretrained 2D Resnet classification network is used to extract a feature vector out of every slice image, then all the features are pooled using max-pooling. This feature is sent to a few FC layers to make the final classification prediction.  In \cite{CTCAPS}, a Capsule network is used to extract feature vector for every image, then these feature vectors are pooled using max-pooling into a global feature vector and a decision is made for the volume. It is claimed that this method is good for small training dataset.  In \cite{SPGC2}, a feature vector is extracted for every image, then multiple pooling methods are ensembled to generate a global feature vector before a final classification is made. In \cite{kollias2021mia}, an RNN is used to aggregate 2D features, but the performance is poor.   

In the 3D CNN method, in \cite{Tongji}, a 3D CNN network is used with both the slice image and a segmented lung mask as input. They discard a fixed percentage of slice images at the beginning and end of a CT-scan volume. In \cite{SPGC3}, the authors first segment the lung mask from a slice image using traditional morphological transforms, then use this mask to select good slice images and generate lung-only images. To make the number of images a fixed number, they use a 3D cubic interpolation to regenerate slice images. In \cite{He2021CovidNet3D}, a 3D CNN network using a fixed number of slice images as input is used. Instead of using a fixed 3D CNN architecture, an autoML method is used to search for best 3D CNN architecture in the network space with MobileNetV2 \cite{MobileNetV2} block. In \cite{iccvCovid}, a 3D CNN with BERT is used on selected slice images of a patient. Sampling is used to make the number of slice images a fixed number.

\section{Data Augmentation and Processing}

In this section, we discuss all the data preparation and augmentation techniques we explore. This data augmentation is an unique one crucial to our final performance. It applies only to CT images where a HU normalization is needed.   

\subsection{DICOM to PNG Conversion}

The dataset is given in DICOM format and the CT image is in Hounsfield unit (HU). So the first thing is to convert the DICOM format to image format. Most of the public datasets are in PNG or JPG format, except for \cite{COVID-CTSet} which uses a TIFF format. We use the PNG format in order to leverage other datasets in PNG format.  

We follow the Kaggle tutorial \cite{Kaggle-Dicom-Tutorial} to convert the DICOM to PNG format. After reading the DICOM file, we extract the slice thickness, slope and intercept. The HU value is, 

\begin{equation}
    HU = slope * DICOM + intercept 
    \label{eq1}
\end{equation}

As explained in the tutorial, different HU values correspond to different materials in human's body, and background. The very important step is to do the HU normalization, from HU value to the PNG value as follows,  

\begin{equation}
    PNG = crop(\frac{HU- HU_{min}}{HU_{max} - HU_{min}},0,1)*255
    \label{eq2}
\end{equation}

where $clip$ is a function to limit the value in range [0,1], $HU_{max}$ and $HU_{max}$ are the maximum and minimum HU value for normalization. These pair ($HU_{min}$, $HU_{max}$) is called a HU window.  

\subsection{HU Augmentation}

\begin{figure}[t]
    \centering
    \includegraphics[scale=0.5]{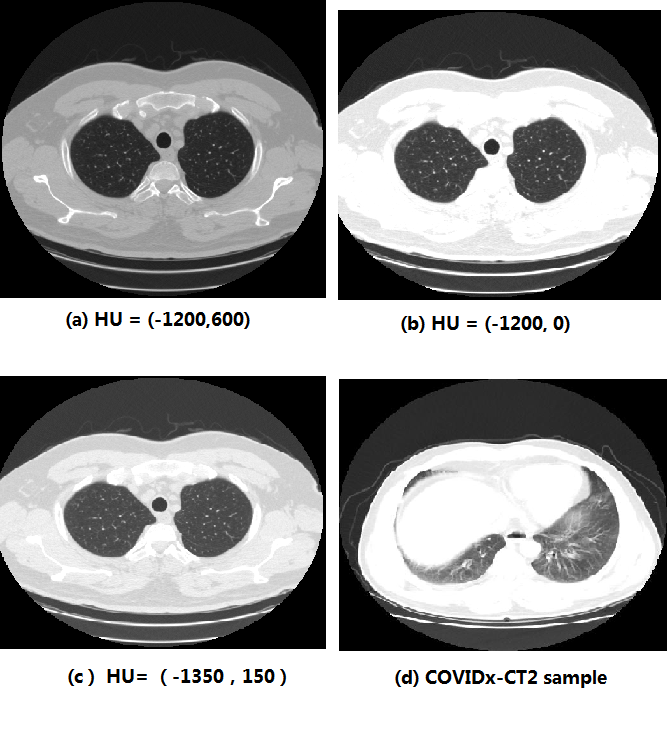}
    \caption{Sample images with three HU windows. }
    \label{fig1}
\end{figure}

\begin{figure}[t]
    \centering
    \includegraphics[scale=0.5]{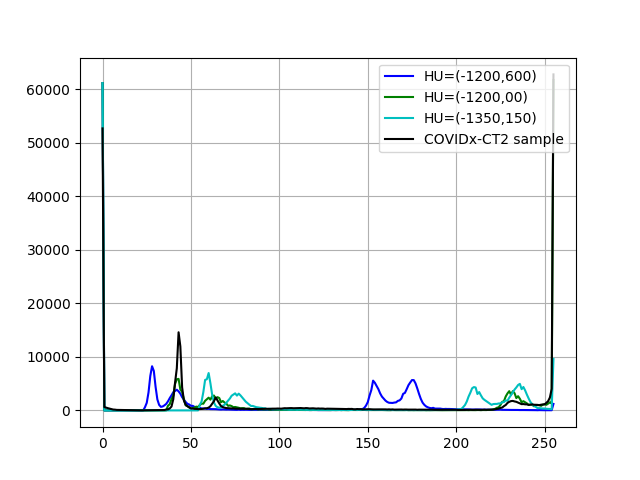}
    \caption{Histogram of the sample images.}
    \label{fig2}
\end{figure}

Even though there are a lot of public CT image dataset, unfortunately, we cannot find anywhere what HU window should be used in the DICOM to PNG conversion. The COVIDx-CT-2 \cite{Gunraj2021} uses [-1350, 150] as default value, the \cite{Tongji} uses [-1200, 600]. We test different HU window in our study and find that they give very similar classification results. However, when we want to use other public dataset, our HU window has to match that of the dataset in order to have a reasonable result. We analyze the image intensity histogram of the COVIDx-CT-2 \cite{Gunraj2021}, and find that if we use a HU window [-1200, 0], the SPGC image intensity histogram can match closely that of the COVIDx-CT-2. Shown in Fig. \ref{fig1} are sample images using these three HU windows and a sample from COVIDx-CT-2. We notice that the Fig. \ref{fig1} (a) is more natural in human eyes, but it may not be the best for CNN classification. Shown in Fig. \ref{fig2} are the histograms of these four sample images. We see that the one using the HU window [-1200,0] matches that of the COVIDx-CT-2 sample well.  

To solve this problem, we propose the so called HU augmentation - to use multiple HU windows in the HU normalization, 

\begin{equation}
    PNG^i = crop(\frac{HU- HU^i_{min}}{HU^i_{max} - HU^i_{min}},0,1)*255, i=1,2,3,...
    \label{eq2_2}
\end{equation}
So the PNG datasets consist of all PNG images from $PNG^1,PNG^2,PNG^3,...$. This not only may overcome the HU window mismatch problem, but also provides more data for training the classification network. We find in the final benchmarking that this is a crucial contribution to our results due to the improved generalization.  

Furthermore, this provides us with another ensemble - the ensemble of test data for prediction. If we use three HU windows to prepare the data, we have three images for every original DICOM slice image, we can have three predictions and we can post process them to get the best performance. we will show these effects in our experiments.

\subsection{Cropping Off Background}

We find that the useless background in the CT image interferes with the training and prediction of the classification network. Therefore, we use lung segmentation mask to cut of the background and only keep the useful portion. This lung segmentation uses traditional image morphological transforms, similar to the Kaggle tutorial \cite{Kaggle-Dicom-Tutorial}. We also train a simple object detection network and its performance is about same as the morphological transform.  This process is shown in Fig. \ref{fig3}(b).  

\subsection{Filtering Closed-Lung Images}

We use an idea similar to \cite{COVID-CTSet} to filter out closed lung images. We use two metrics - the percent of segmented lung in the total image, and the one in \cite{COVID-CTSet}. Please note that in \cite{COVID-CTSet}, the images are not cropped, so the filter does not work consistently. We use the idea on the cropped images, the so region of interest (ROI) is aligned more consistently on different CT images. 

\begin{figure}[t]
    \centering
    \includegraphics[scale=0.5]{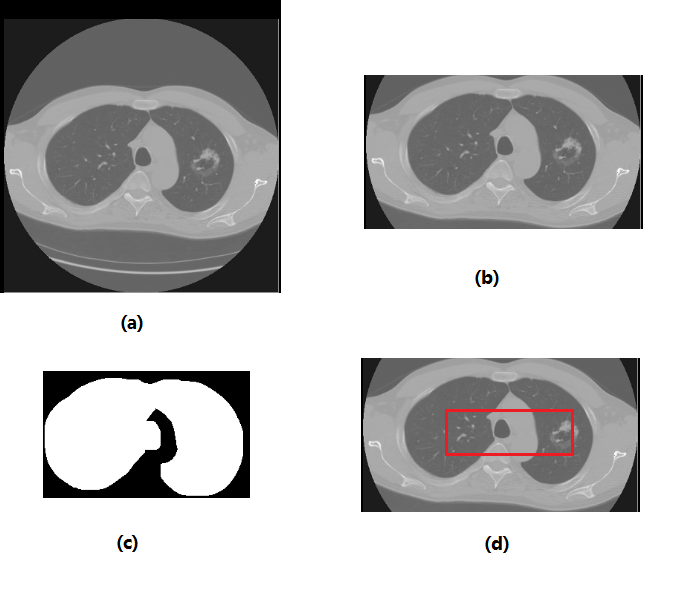}
    \caption{Processing of CT-scan image, (a) an original image with a circle background mask, (b) after cropping off background, (c) the extracted lung mask, (d) the closed-lung filtering window.}
    \label{fig3}
\end{figure}

In the first metric, we keep slice images whose percent of segmented lung is more than 10\% of the total image.  Shown in Fig. \ref{fig3}(c) is an extracted lung mask. When the percentage of the white pixels is less than $10\%$ of the total pixels, the lung is closed.     
In the second metric, we count the number of black pixels (intensity value $<100$) in the ROI = ([120,240], [370,340]) in a slice image, as shown in the red rectangle in Fig.\ref{fig3}(d). Then among all the slice images of a patient, find the maximum and minimum values, and use a threshold = (max-min)/factor, where factor is typically 1.5-3. Slice images whose number of black pixels in the ROI is less than the threshold are filtered out \cite{COVID-CTSet}.

\subsection{Leveraging other Datasets}

Even with the HU augmentation, the training dataset is still very small. We explore the CT image dataset available in the literature. There are quite a lot of them, we namely some of them here COVIDx-CT-2 \cite{Gunraj2021}, CNCB \cite{CNCB}, CT-Codeset \cite{COVID-CTSet}. Out of all these datasets, COVIDx-CT-2 \cite{Gunraj2021} is the largest one with nearly 200K images, we decide to use it in our study. 

However, we do not want these third-party dataset to overrun the SPGC dataset. So we only use a small portion of it, so the total number from it is not more than that in the SPGC dataset. We use this dataset in both the training and validation dataset, but not in the test dataset.  We only use the SPGC images in the test dataset.   

\section{Classification Network}

\begin{figure}[t]
    \centering
    \includegraphics[scale=0.3]{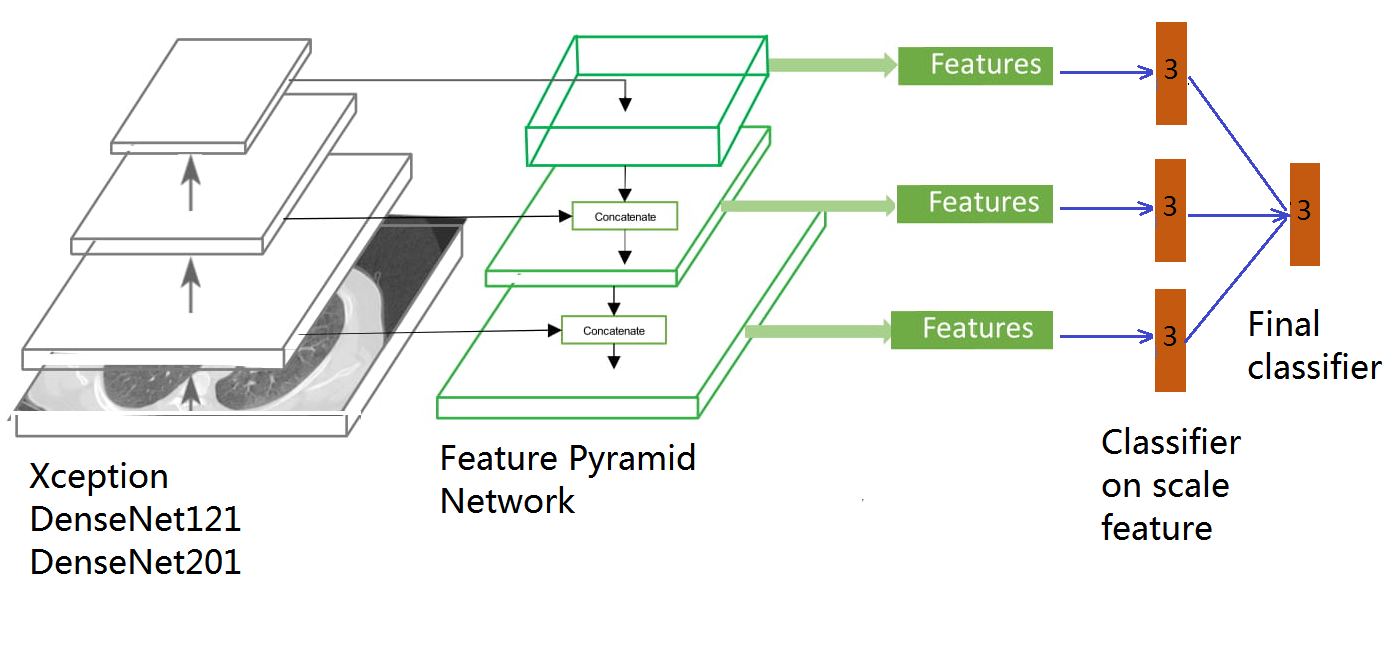}
    \caption{CNN-FPN classifier for COVID-19, CAP and normal case.}
    \label{fig4}
\end{figure}

Our 2D CNN classification network is shown in Fig. \ref{fig4}. We use a network similar to \cite{COVID-CTSet}. Three different scale features are generated in the FPN, and a three-class classifier is applied on every feature. At the end, these three classifiers are merged into the final three-class classifier.

Some details of our implementation are listed here. We use a three-way classification, no matter the test dataset is a two-way or three-way classification task. We use rotation, shift, and scale transforms. We find that the rotation degree has big impact on the classification performance. We limit it to 15 degree. We use batch normalization after all convolutional layers. We use the image size 224x224 in ResNet50 \cite{Resnet} and DenseNet121/201 \cite{DenseNet}, and 299x299 in Xception \cite{Xception}.

We test different networks, including ResNet50 \cite{Resnet}, MobileNetV2 \cite{MobileNetV2}, Xception \cite{Xception}, DenseNet121 and DenseNet201 \cite{DenseNet}. We find that the DenseNet trains faster and achieves good performance. So in our ablation study, we use the DenseNet201. In the final classification benchmarking, we use an ensemble of the Xception, DenseNet121, and DenseNet201. In order not to use too much CPU or GPU memory at a time, we run the three models one by one and post-process the results.

We also test a regular 2D CNN classification network without FPN, such as a ResNet50, as well as a small private CNN, we find that they can achieve good accuracy with all our other data augmentation and training techniques. However, the CNN with FPN can achieve a better accuracy, usually $1-2\%$ better than the network without FPN.

\section{Experiments}

We use Keras-2.3.0 and Tensorflow-GPU-2.2 in our implementation. We use learning rate 1E-4 at the beginning then adjust to 1E-5. We use typically 50 epochs to train the CNN network, except for fine tuning where a smaller number of epochs is used.  A class weight is used when the numbers of images for the three classes - COVID, CAP, normal are unbalanced.    

\subsection{Dataset}

The dataset provided by SPGC \cite{SPGC} includes 307 patients, which are diagnosed by medical experts. Out of these patients, 171 patients have COVID-19, 60 patients have CAP, and the rest 76 are normal cases. This dataset is not small, however, only a small portion including 55 COVID-19 patients and 25 CAP patients has slice annotations. This limited slice annotation turns out a big challenge in the slice based classification.

Since we use a slice image based classification model, we use all the annotated slice images as the training and validation datasets. For COVID and normal cases, we use a 7:3 split ratio for training and validation dataset, and use a 9:1 split ratio for the CAP cases since there are a lot less of them. All COVID/CAP patients in the validation dataset are not used in the training, only a small portion of slice images of the normal patients are used in the training. In the patient-wise classification, we use all the patients without slice annotation as the validation dataset.

Furthermore, we leverage the CT image dataset we can find online - the COVIDx-CT-2 dataset \cite{Gunraj2021}. The combined dataset makes our trained model generalize much better than using one single HU window alone.

\subsection{From Slice Classification to Patient Classification}
The slice based classification network predicts a result for every slice image. Our goal, however, is to have a classification result on patient. Given the number of slice images per patient, we use two metrics to make the final decision. In the first method, we use a slice threshold $th_s$,     
\begin{equation}
\begin{aligned}
& if \; n(COVID) > n(CAP) >= th_s, \; decision = COVID    \\
& else \; if \; n(CAP) > n(COVID) >= th_s, \; decision = CAP   \\ 
& else \; if \; n(COVID) >= th_s, \; decision = COVID  \\ 
& else \; if \; n(CAP) >= th_s, \; decision = CAP  \\ 
& else, \; decision = Normal \\
\end{aligned}
\end{equation}
where $n()$ is the number of slice classification. We can change this rule to use the percent threshold $th_p$, where the $n()$ is replaced by a percent $n()$ - the ratio of the number of slice classification of each class out of the total slice number of every patient.   

Please note that, in our final decision, we do not do majority vote on multiple model predictions on every image. Instead, we mix all image predictions from all three models and make a final decision based on the number of predictions be-
longing to the three classes. The performance of this rule is noticeably better than the majority vote on a single image. 

At validation/test time, we first do slice image classification, then use the above rule to make decision for a patient. We use the un-annotated patients as the validation patient dataset. We optimize the parameters of our classifier on the validation dataset, then use the optimized parameters on the final test datasets for the challenge submission.

\subsection{HU Augmentation}
The use of HU augmentation is our unique novelty. So we do some ablation study to demonstrate its effects using the DenseNet201-FPN network. In this experiment we use the factor = 1.5 in the closed-lung filtering.  We use three HU windows: SPGC3 = [-1000,400], SPGC4 = [-1200,0], SPGC6 = [-1200,600]. In Table 1, we list the result of individual HU window in the first panel, and list the result of the HU augmentation (three HU windows combined) in the second panel.  

\begin{table}
	\begin{center}
		\caption{Patient classification using HU augmentation. The DenseNet201 is used}
		\label{T1}
		\begin{tabular}{ccc}
		    \hline
			Train Data &  Test Data & Accuracy  \\
			\hline
			SPGC3 & SPGC3 & 0.8138 \\
			SPGC4 & SPGC4 & 0.8105 \\
			SPGC6 & SPGC6 & 0.7797 \\
			\hline
			SPGC3+4+6 & SPGC3 & 0.8194\\
			SPGC3+4+6 & SPGC4 & 0.8194\\
			SPGC3+4+6 & SPGC6 & 0.8194\\
			\hline
			+COVIDx CT-2 & SPGC3 & \textbf{0.9075}\\
			+COVIDx CT-2 & SPGC4 & 0.9031\\
			+COVIDx CT-2 & SPGC6 & 0.8458\\
			\hline
		\end{tabular}
	\end{center}
\end{table}

From the results in the first panel, we notice that the SPGC3 and SPGC4 are two good HU windows, while SPGC6 is not a good one.  From the results in the second panel, the test results of three HU windows are about the same, even though their sensitivities (not shown) are not the same. Even the individually bad HU window (SPGC6) now has a good performance. 

\subsection{Leveraging Other Dataset}

In addition to the HU augmentation, we use the COVIDx-CT-2 dataset \cite{Gunraj2021} in our training. As we mentioned before, we only use a small portion of it, so it does not overrun the SPGC dataset. So in our training dataset, we include both the three HU window augmented SPGC dataset, and the selected COVIDx-CT-2 dataset. The patient classification results are listed in the third panel of Table \ref{T1}.

From the results, we notice that the addition of the extra training data has a big boost to the patient classification accuracy, from previous about $82\%$ to nearly $91\%$. On the other hand, SPGC6 is a lot worse than the SPGC3 and SPGC4, even though its accuracy is improved from without using COVIDx-CT-2 dataset. Based on the poor performance of SPGC6, we do not use it in the ensemble of HU windows in the final tests.

\subsection{Ensemble of Classification Models and HU Windows}

We test a few classification networks including ResNet50, MobileNetV2, and others. Based on the patient classification accuracy results, we choose to use the Xception, DenseNet101 and DenseNet201. The individual model results are listed in first panel of Table \ref{T2}.

\begin{table}
	\begin{center}
		\caption{Patient classification accuracy and sensitivity (COVID,CAP,Normal) on the training/validation dataset.}
		\label{T2}
		\begin{tabular}{cccc}
		    \hline
			Network & Test Data & Accuracy & Sensitivity\\
			\hline
			DenseNet201 & SPGC3 & 0.9031 & (0.8,0.871,1.0)\\
			DenseNet201 & SPGC4 & 0.8942 & (0.771,0.862,1.0) \\
			DenseNet121 & SPGC3 & 0.9031 & (0.8,0.871,1.0) \\
			DenseNet121 & SPGC4 & 0.9031 & (0.771,0.879,1.0]) \\
			Xception & SPGC3 & 0.8942 & (0.714,0.888,0.987)\\
			Xception & SPGC4 & 0.9163 & (0.8,0.897,1.0) \\
			\hline
			Ensemble & SPGC3 & 0.9295 & (0.8,0.922,1.0)\\
			Ensemble & SPGC4 & 0.9251 & (0.8,0.905,1.0)\\
			\hline
			Ensemble & SPGC3+4 & \textbf{0.9339} & (0.8,0.931,1.0)\\
			\hline			
		\end{tabular}
	\end{center}
\end{table}

The three networks give similar patient classification accuracy results on the two HU window datasets SPGC3 and SPGC4. The DenseNet201 has the best accuracy on SPGC3, and the Xception has best accuracy on SPGC4. Adding all three network into an ensemble, and on the ensemble of SPGC3 and SPGC4, our final accuracy result on the SPGC training/validation dataset is 93.39\%.      

\subsection{Results on the Test Dataset}

On the new test dataset that may come from a different domain, we can tune the thresholds $th_s$ and $th_p$ on a small portion of the dateset to achieve good performance, then use the two thresholds on the rest of the data. Listed in Table \ref{T3} are the results. Since the ground truth of this dataset has not been released, we cannot optimize the parameters $th_s$ and $th_p$ other than using the optimized values on the validation dataset.

\begin{table}
	\begin{center}
		\caption{PPatient classification accuracy and sensitivity (COVID,CAP,Normal) on the test dataset.}
		\label{T3}
		\begin{tabular}{ccc}
		    \hline
			($th_s$,$th_p$) & Accuracy & Sensitivity\\
			\hline
			(2,1\%) & 0.6778 & (0.8857,1.0,0.2857)\\
			(5,5\%) & 0.8111 & (0.8286,1.0,0.6857)\\
			\hline			
		\end{tabular}
	\end{center}
\end{table}

\section{Conclusions}

We provide a solution for the COVID-19 automatic diagnosis on the SPGC dataset. The key novelty is a data augmentation using multiple HU normalization windows. With all techniques put together, we achieve a patient classification accuracy $93.39\%$ on the provided training/validation dataset, and an accuracy at least $81.11\%$ on the test dataset.

\bibliographystyle{IEEEbib}
\bibliography{strings,refs}

\begin{thebibliography}{10}

\bibitem{WHO}
``Coronavirus disease (covid-19) pandemic,''
\newblock {\em
  https://www.who.int/emergencies/diseases/novel-coronavirus-2019}.

\bibitem{COVID-trace}
W.~Tan and J.~Liu,
\newblock ``Application of face recognition in tracing covid-19 patients and
  close contacts,''
\newblock {\em IEEE ICMLA}, 2020.

\bibitem{SPGC}
``2021 ieee icassp signal processing grand challenge (spgc) covid-19
  radiomics,''
\newblock {\em https://2021.ieeeicassp.org/GrandChallenge.asp/CoVID}, 2021.

\bibitem{Hybrid-COVID}
K.~Bayoudh, F.~Hamdaoui, and A.~Mtibaa,
\newblock ``Hybrid-covid: a novel hybrid 2d/3d cnn based on cross-domain
  adaptation approach for covid-19 screening from chest x-ray images,''
\newblock {\em Physical and Engineering Sciences in Medicine volume}, vol. 43,
  2020.

\bibitem{SODA}
J.~Zhou, B.~Jing, and Z.~Wang,
\newblock ``Soda: Detecting covid-19 in chest x-rays with semi-supervised open
  set domain adaptation,''
\newblock {\em Arxiv preprint}, 2020.

\bibitem{Contrative-CovidNet}
Z.~Wang, Q.~Liu, and Q.~Dou,
\newblock ``Contrastive cross-site learning with redesigned net for covid-19 ct
  classification,''
\newblock {\em Published in: IEEE Journal of Biomedical and Health
  Informatics}, vol. 24, 2020.

\bibitem{Covid-caps}
P.~Afshar, S.~Heidarian, F.~Naderkhani, A.~Oikonomou, K.~Plataniotis, and
  A.~Mohammadi,
\newblock ``Covid-caps: A capsule networkbased framework for identification of
  covid-19 cases from c-ray images,''
\newblock {\em Pattern Recognition Letters}, vol. 138, 2020.

\bibitem{COVID-FACT}
S.~Heidarian, P.~Afshar, N.~Enshaei, F.~Naderkhani, A.~Oikonomou, F.~B. Fard,
  K.~Samimi, K.N. Plataniotis, A.~Mohammadi, and M.J. Rafiee,
\newblock ``Covid-fact: A fully-automated capsule network-based framework for
  identification of covid-19 cases from chest ct scans,''
\newblock {\em Arxiv preprint 2010.16041}, 2020.

\bibitem{DenseNet}
G.~Huang, Z.~Liu, L.~Maaten, and K.Q. Weinberger,
\newblock ``Densely connected convolutional networks,''
\newblock {\em CVPR}, 2017.

\bibitem{Xception}
Chollet F.,
\newblock ``Xception: Deep learning with depthwise separable convolutions,''
\newblock {\em CVPR}, 2017.

\bibitem{FPN}
T~Lin, P.~Dollár, R.~Girshick, K.~He, B~Hariharan, and S.~Belongie,
\newblock ``Feature pyramid networks for object detection,''
\newblock {\em CVPR}, 2017.

\bibitem{example1}
S.~Sabour, N~Frosst, and G.~Hinton,
\newblock ``Deep learning approaches for covid-19 detection based on chest
  x-ray images,''
\newblock {\em Expert Systems with Applications}, vol. 164, 2021.

\bibitem{example2}
A.~Amyar, R.~Modzelewski, H.~Li, and S.~Ruan,
\newblock ``Multi-task deep learning based ct imaging analysis for covid-19
  pneumonia: Classification and segmentation,''
\newblock {\em Computers in Biology and Medicine}, vol. 126, 2020.

\bibitem{Review1}
I.~Ozsahin, B.~Sekeroglu, M.~Musa, M.~Mustapha, and D~Ozsahin,
\newblock ``Review on diagnosis of covid-19 from chest ct images using
  artificial intelligence,''
\newblock {\em Computational and Mathematical Methods}, vol. 2020, 2020.

\bibitem{Review2}
O.S Albahri and et~al.,
\newblock ``Systematic review of artificial intelligence techniques in the
  detection and classification of covid-19 medical images in terms of
  evaluation and benchmarking: Taxonomy analysis, challenges, future solutions
  and methodological aspects,''
\newblock {\em J Infect Public Health}, p. 1381–1396, 2020.

\bibitem{SPGC1}
Shubham Chaudhary, Sadbhawna, Vinit Jakhetiya, Badri Subudhi, Ujjwal Baid, and
  Sharath Guntuku,
\newblock ``Detecting covid-19 and community acquired pneumonia using chest ct
  scan images with deep learning,''
\newblock in {\em ICASSP}, 2021.

\bibitem{COVID-CTSet}
M.~Rahimzadeh, A.~Attar, and S.~M. Sakhaei,
\newblock ``A fully automated deep learning-based network for detecting
  covid-19 from a new and large lung ct scan dataset,''
\newblock {\em medRxiv}, 2020.

\bibitem{li2020artificial}
Lin Li, Lixin Qin, Zeguo Xu, Youbing Yin, Xin Wang, Bin Kong, Junjie Bai,
  Yi~Lu, Zhenghan Fang, Qi~Song, Kunlin Cao, et~al.,
\newblock ``Artificial intelligence distinguishes covid-19 from community
  acquired pneumonia on chest ct,''
\newblock {\em Radiology}, 2020.

\bibitem{CTCAPS}
Shahin Heidarian, Parnian Afshar, Arash Mohammadi, Javad Rafiee, Anastasia
  Oikonomou, Konstantinos Plataniotis, and Farnoosh Naderkhani,
\newblock ``Ct-caps: Feature extraction-based automated framework for covid-19
  disease identification from chest ct scans using capsule networks,''
\newblock in {\em ICASSP}, 2021, pp. 1040--1044.

\bibitem{SPGC2}
Pratyush Garg, Rishabh Ranjan, Kamini Upadhyay, Monika Agrawal, and Desh
  Deepak,
\newblock ``Multi-scale residual network for covid-19 diagnosis using
  ct-scans,''
\newblock in {\em ICASSP}, 2021, pp. 8558--8562.

\bibitem{Tongji}
X.~Wang, X.~Deng, Q.~Fu, Q.~Zhou, J.~Feng, H.~Ma, W.~Liu, and Q.~Zheng,
\newblock ``A weakly-supervised framework for covid-19 classification and
  lesion localization from chest ct,''
\newblock {\em IEEE Transactions on Medical Imaging}, vol. 39, pp. 2615--2625,
  August 2020.

\bibitem{SPGC3}
Shuohan Xue and Charith Abhayaratne,
\newblock ``Covid-19 diagnostic using 3d deep transfer learning for
  classification of volumetric computerised tomography chest scans,''
\newblock in {\em ICASSP}, 2021.

\bibitem{He2021CovidNet3D}
Xin He, Shihao Wang, Xiaowen Chu, Shaohuai Shi, Jiangping Tang, Xin Liu,
  Chenggang Yan, Jiyong Zhang, and Guiguang Ding,
\newblock ``Automated model design and benchmarking of 3d deep learning models
  for covid-19 detection with chest ct scans,''
\newblock {\em Proceedings of the AAAI Conference on Artificial Intelligence},
  2021.

\bibitem{COVID-CT-Mask-Net}
Aram Ter-Sarkisov,
\newblock ``Covid-ct-mask-net: Prediction of covid-19 from ct scans using
  regional features,''
\newblock {\em medRxiv}, 2020.

\bibitem{Gunraj2020}
H.~Gunraj, L.~Wang, and A.~Wong,
\newblock ``Covidnet-ct: A tailored deep convolutional neural network design
  for detection of covid-19 cases from chest ct images,''
\newblock {\em Frontiers in Medicine}, vol. 7, pp. 1025, 2020.

\bibitem{Gunraj2021}
H.~Gunraj, A.~Sabri, D.~Koff, and A.~Wong,
\newblock ``Covid-net ct-2: Enhanced deep neural networks for detection of
  covid-19 from chest ct images through bigger, more diverse learning,'' 2021.

\bibitem{kollias2021mia}
D.~Kollias, A.~Arsenos, L.~Soukissian, and S.~Kollias,
\newblock ``Mia-cov19d: Covid-19 detection through 3-d chest ct image
  analysis,''
\newblock {\em arXiv preprint arXiv:2106.07524}, 2021.

\bibitem{MobileNetV2}
M~Sandler, A.~Howard, M.~Zhu, A.~Zhmoginov, and L.~Chen,
\newblock ``Mobilenetv2: Inverted residuals and linear bottlenecks,''
\newblock {\em CVPR}, 2018.

\bibitem{iccvCovid}
W.~Tan and J.~Liu,
\newblock ``A 3d cnn network with bert for automatic covid-19 diagnosis from
  ct-scan images,''
\newblock {\em ICCV Workshops}, 2021.

\bibitem{Kaggle-Dicom-Tutorial}
``full-preprocessing-tutorial,''
\newblock {\em https://www.kaggle.com/gzuidhof/full-preprocessing-tutorial}.

\bibitem{CNCB}
K.~Zhang, X.~Liu, J.~Shen, and et~al.,
\newblock ``Clinically applicable ai system for accurate diagnosis,
  quantitative measurements and prognosis of covid-19 pneumonia using computed
  tomography,''
\newblock {\em Cell}, April 2020.

\bibitem{Resnet}
K.~He, X.~Zhang, S.~Ren, and J.~Sun,
\newblock ``Deep residual learning for image recognition,''
\newblock {\em CVPR}, 2015.

\end{thebibliography}

\end{document}